\documentclass[aps,prl,twocolumn,showpacs,superscriptaddress,letterpaper,floatfix]{revtex4}

\usepackage{graphicx}
\usepackage{longtable}
\usepackage{amssymb}
\usepackage{amsthm}
\usepackage{amsfonts}
\usepackage[centertags]{amsmath}
\usepackage{color}
\usepackage[latin1]{inputenc}
\usepackage[english]{babel}
\usepackage{xspace}
\usepackage{verbatim}

\newcommand{\lag}{LG$_{33}$\xspace}
\newcommand{\lgzero}{LG$_{00}$\xspace}

\begin{document}
\title[Generation of high-purity higher-order Laguerre-Gauss beams at high laser power]
{Generation of high-purity higher-order Laguerre-Gauss beams at high laser power}

\author{L. Carbone}
\email{lc@star.sr.bham.ac.uk}
\address{School of Physics and Astronomy, University of
Birmingham, Edgbaston, Birmingham B15 2TT, UK}
\author{C. Bogan}
\address{Max Planck Institute for Gravitational Physics (Albert Einstein Institute) and Leibniz Universit\"at Hannover, Callinstra{\ss}e 38, 30167 Hannover, Germany}
\author{P. Fulda}
\address{School of Physics and Astronomy, University of
Birmingham, Edgbaston, Birmingham B15 2TT, UK}
\author{A. Freise}
\address{School of Physics and Astronomy, University of
Birmingham, Edgbaston, Birmingham B15 2TT, UK}
\author{B. Willke}
\address{Max Planck Institute for Gravitational Physics (Albert Einstein Institute) and Leibniz Universit\"at Hannover, Callinstra{\ss}e 38, 30167 Hannover, Germany}

\begin{abstract} 	
We have investigated the generation of highly pure higher-order Laguerre-Gauss (LG)
beams at high laser power of order 100\,W, the same regime that will be used by 
2$^{\rm nd}$ generation gravitational wave interferometers such as Advanced LIGO. 
We report on the generation of a helical type \lag mode with a
purity of order 97\,\% at a power of 83\,W, the highest power ever reported in literature
for a 
higher-order LG mode. 
\end{abstract}
\date{\today}

%
\maketitle

\noindent{\bf Introduction} 
The generation of Laguerre Gauss (LG) optical beams has 
significantly gained interest in recent times. 
LG modes present in fact several unusual features that make 
them suitable for a wide range of applications. 
In physics for example, {\it donut-shaped} LG beams 
confine particles in optical traps \cite{lg:mot,lg:dipole} 
or speed-up charged particles in particle accelerators \cite{lg:acc};
higher-order {\it multi-ringed} LG beams form 
toroidal traps for Bose-Einstein condensates \cite{lg:bec};
LG beams act as {\it optical spanners} transferring 
their orbital angular momentum to spin macroscopic particles \cite{lg:span}. 
In the last years, use of LG beams has been 
reported in the most diverse areas of science, 
some examples are material processing \cite{lg:matproc}, 
microscopy \cite{lg:micro}, 
lithography \cite{lg:lith}, 
motion sensors \cite{lg:sens},
biology \cite{lg:biol},
biomedics \cite{lg:biom}. 

Higher-order helical type LG modes have been also proposed as upgrades to 
the readout beams of  2$^{\rm nd}$ generation gravitational wave (GW) interferometers
such as Advanced LIGO \cite{aligo} and Advanced VIRGO \cite{avirgo}, 
and are currently baselined for the Einstein Telescope \cite{ET}.
The wider, more uniform transverse intensity distribution of a subset
of these beams, compared to the currently used \lgzero fundamental mode, 
can effectively average over the mirror surface fluctuations, 
to mitigate the effects of brownian motion of the 
mirror surfaces
on the detector GW sensitivity \cite{Mours06,Vinet09,Vinet10}. 
Using LG modes can also lead to a reduction of thermal effects 
such as distortions in the mirror substrates, 
when operating at the high laser power regime envisioned for these detectors \cite{Vinet07}.
Theoretical studies have initially proven the compatibility of LG modes 
with the control schemes commonly employed, 
and identified the \lag beams as a good trade-off between mirror thermal noise suppression 
and beam clipping losses \cite{Chelkowski:PRD}. 
Subsequent laboratory experiments have then 
demonstrated the generation of LG modes 
at the required purity, and the possibility of implementing interferometric measurements 
using LG beams \cite{Fulda:PRD,granata}. 

One crucial step into a realistic implementation of LG modes in GW interferometers is 
to demonstrate the generation of such beams at the high power levels of order 100\,W 
foreseen by next generation detectors. 
High power LG beams should also comply with the stringent requirements that current 
GW laser sources have successfully achieved, and present comparably high levels of purity, 
stability and low noise \cite{ref:aLIGO:PSL,ref:aLIGO:PSL2,ref:aLIGO:PSL3}.
Generation of LG beams of order tens of W has been reported in literature 
\cite{lg_high:1,lg_high:2,lg_high:3,lg_high:4},
although for different types of applications and 
limited to lower order {\it donut-shape} LG$_{01}$ beams only. 
These beams do not meet the requirements discussed above and furthermore 
are based on beam shaping techniques which have little adaptability and are 
hardly exportable to higher-order modes or to more generic applications.

We have investigated the generation of higher-order LG modes at the high laser power regime 
required for operating 2$^{\rm nd}$ generation GW interferometers at full sensitivity. 
The experiment is based on a beam preparation method originally developed 
by some of the authors at low power \cite{Fulda:PRD} and potentially scalable to full scale interferometers. 
Our investigation aimed not only to generate higher-order LG beam at the highest 
possible laser power, mode purity and conversion efficiency,
but also to identify potential limits of the technology. 
In this letter we present our experiment's details and discuss the results. 
We stress that, due to the simplicity of the experimental scheme, 
this method is adaptable to a variety of 
applications, therefore the presented 
results are potentially relevant to a broader audience than the GW community. 

\noindent
 {\bf LG modes:} 
LG modes are a complete and orthogonal set of solutions for the paraxial wave equation. 
The complex amplitude of a helical type LG$_{pl}$ mode, 
with radial and azimuthal indices $p$ and $l$, is usually described as~\cite{Lasers}:
\begin{eqnarray}    
\label{eq:1}
{\rm LG}_{pl}^{\rm hel}\left( r,\phi,z \right )&= & 
		 \frac{1}{w(z)}
		 \sqrt{\frac{2p!}{\pi(|l|+p)!}} 
		 \left(\frac{\sqrt{2}r}{w(z)}\right)^{|l|} 
		 L^{|l|}_{p}\left(\frac{2r^{2}}{w^{2}(z)}\right) 
		 \nonumber \\
		 &\times & e^{i\left(2p+|l|+1\right)\Psi(z)}
		 e^{\left(-\frac{ikr^{2}}{2R_{c}(z)}-\frac{r^{2}}{w^{2}(z)}+il\phi\right)}.
\end{eqnarray}
Here $(r,\theta,z)$ are cylindrical-polar coordinates, $k$ is the wavenumber, $w(z)$ the beam radius, 
$R_{c}(z)$ the radius of curvature of the beam wavefront, $\Psi(z)$ the Gouy phase, 
and $L^{|l|}_{p}(x)$ are the generalised Laguerre polynomials. 
LG beams are axisymmetric and have spherical wavefront, 
so they are natural eigenmodes of optical systems whose 
optical surfaces are spherical and whose symmetry is cylindrical.
The order of a LG mode is given by the number $\left(2p+|l|\right)$:
when circulating in optical resonators, modes of the same order experience same 
resonance conditions, due to the $e^{\left[i\left(2p+|l|+1\right)\Psi(z)\right]}$ phase term, 
so the cavity is degenerate for this family of modes.
The $L^{|l|}_{p}(x)$ term is what gives LG modes their characteristic {\it ringed} shape,
while the azimuthal phase dependence $e^{il\phi}$ is responsible 
for their orbital angular momentum, $l\hbar$ per photon.

\begin{figure}[t]
\begin{center}
\includegraphics[width=8.5cm]{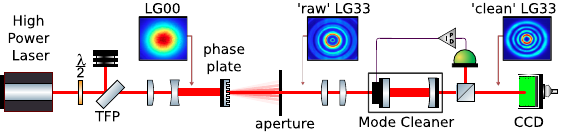}
\caption{\label{fig:exper_setup} 
Cartoon of the experimental setup described in the paper.
The main components are labelled. 
Beam dumps, steering mirrors, 
waveplates and polarisers 
are not shown.}
\end{center}
\end{figure}

\noindent {\bf The experiment} A sketch of the experimental setup is presented in Fig.\,\ref{fig:exper_setup}.
A high power, ideally pure fundamental mode laser beam is 
mode-matched to a desired waist size via a telescope, and then
sent on a diffractive phase plate, an etched glass substrate with 
varying thickness which can imprint the \lag spiralling phase pattern 
onto the wavefront of the input beam.
The diffraction orders are separated with an aperture, 
and the main diffracted beam,
a composite beam with a dominant \lag over a background of higher-order modes of minor intensity, 
is then injected to a linear mode cleaner (MC) cavity, 
which is alternatively used to analyse the mode content of the input beam itself (scan mode)
or to {\it spatially} filter out non order 9 LG modes (locked mode) 
to enhance the mode purity of the \lag beam generated in transmission. 
This is eventually recorded by means of a high dynamic range photodiode, 
for measurement and control purposes, and by a CCD camera, for off-line analysis. 
Light power measurements are performed at different stages along the optical setup, 
namely before and after the phase plate, at the MC input and, 
when the MC cavity is locked, both in reflection and in transmission from the cavity.
Similarly, images of the beam intensity distributions are taken at analogous positions for mode content analyses.

The high-power laser source is the Reference System for the Advanced LIGO pre-stabilised laser (PSL) 
\cite{ref:aLIGO:PSL,ref:aLIGO:PSL2,ref:aLIGO:PSL3} 
and it is located at the Hannover labs where this experiment was performed. 
The PSL consists of a 2\,W Nd:YAG non-planar ring oscillator, 
two stages of amplification (up to 35\,W and 200\,W respectively), 
and a ring cavity at the output, 
which provides filtering for beam's spatial profile, pointing and power fluctuations. 
The output is a {{140\,W}}, 1064\,nm, continuous wave, 99.5\% pure \lgzero beam.

The phase plate mode conversion method \cite{ref:phaseplate} was chosen 
amongst other successful techniques 
\cite{ref:astigmatic,ref:hologram,ref:SLM} 
for the compatibility of passive glass components
with the high power regime to be tested in this experiment,  
and for the relative simplicity of implementation.
Our phase plate is a 3\,mm thick fused silica substrate 
with $3000\times 3000$, 7\,$\mu$m side etched pixels, 
with 8 levels of etching depth resolution 
\footnote{The phase plate was manufactured 
by Jenoptik GmbH based on a custom design by some of the authors \cite{fulda:13}.
}.
The etched grating phase pattern reproduces the spiralling helical \lag mode phase structure. 
On top, a 2.3\,mrad angle blazed grating pattern is superimposed to separate 
the main diffraction order beam from unmodulated residuals of the \lgzero mode \cite{ref:hologram}. 
FFT beam propagation methods and modal analysis of the beam in the far field 
were used to estimate the 
efficiency of this phase plate design in the conversion from {\lgzero} to {\lag},
which is in the region of 75\% depending on the correct size and relative 
alignment of the incident beam with respect to the phase plate itself \cite{fulda:13}.
To avoid having light reflected towards the laser, a 1064nm anti-reflective coating was deposited 
on both surfaces of the phase plate.
Measurements showed that about 95\% of the light power 
successfully transmits into the main diffraction order beam, 
about $4\%$ is dispersed in higher diffraction orders and less than 
0.2\% is reflected towards the laser source.

The MC is a 21\,cm long, plano-concave linear cavity,
with 1" fused silica mirrors glued to the ends of a rigid Al spacer.
Highly-reflective coatings ($R=97.5\%$) were deposited on the 
mirror substrates in a single coating run, aiming for a nominally impedance matched, 
maximised transmission cavity. 
The MC has stability parameter $g \approx 0.8$, Free Spectral Range = 714\,MHz, measured finesse $F\approx 130$. 
Its microscopic length is controlled via a PZT located between the spacer and the input mirror.
The error signal for the feedback control is generated by dithering the input mirror position 
with the PZT and then extracted from the light transmitted by the cavity.

\begin{figure}[t]
\begin{center}
\includegraphics[width=8.5cm]{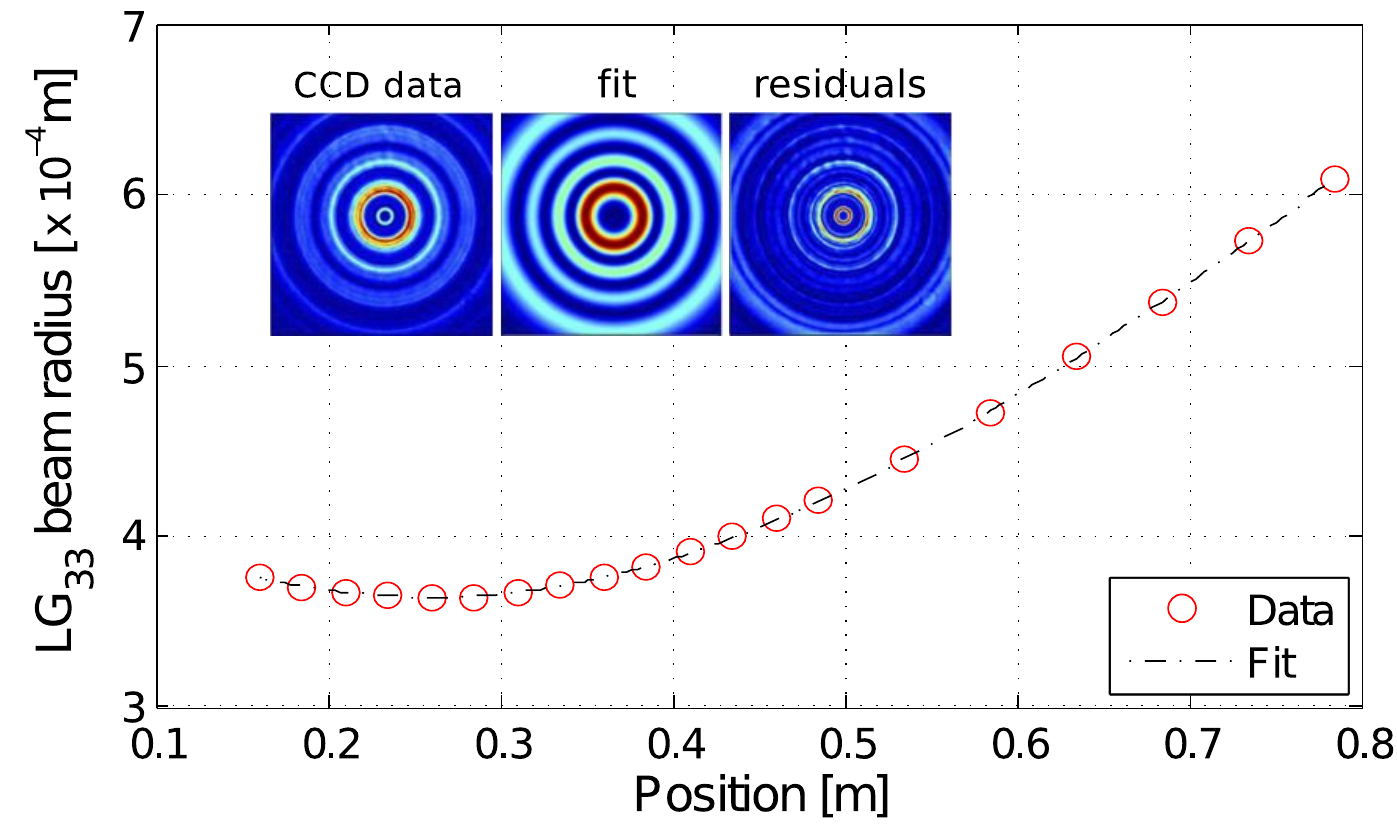}
\caption{\label{fig:lg33_beam_profile} Profile of the \lag beam injected into the MC cavity, 
with best fit shown for comparison.
The insets show an example of fitting of a \lag beam, 
with the intensity patterns of a measured beam compared to the fit and related residuals.
}
\end{center}
\end{figure}
Mode matching of the LG beam generated by the phase plate to the MC eigenmode 
is non trivial and 
proved crucial 
to the successful operation of the cavity. 
Since conventional beam profilers do not usually resolve LG modes,
we first recorded the beam intensity profile with a CCD camera placed 
along the beam path, then 
the images were analysed 
using customised 
fitting scripts \cite{SIMTOOLS} which automatically identify the dominant \lag mode 
and estimate the beam radius at the given position. 
Subsequent adjustments of the lenses rapidly led to match the 
beam waist parameters to within few $\mu$m from the aimed value, 
in this case $w_0$=$365\mu$m. 
In full scale GW interferometers, acceptable matching 
errors are of order 1\%. 
Our result shows that mode matching of higher-order LG beams can be 
performed with comparable 
accuracy. An example of this analysis is given in Fig.\,\ref{fig:lg33_beam_profile}.

\begin{figure}[t]
\begin{center}
\includegraphics[width=8.5cm]{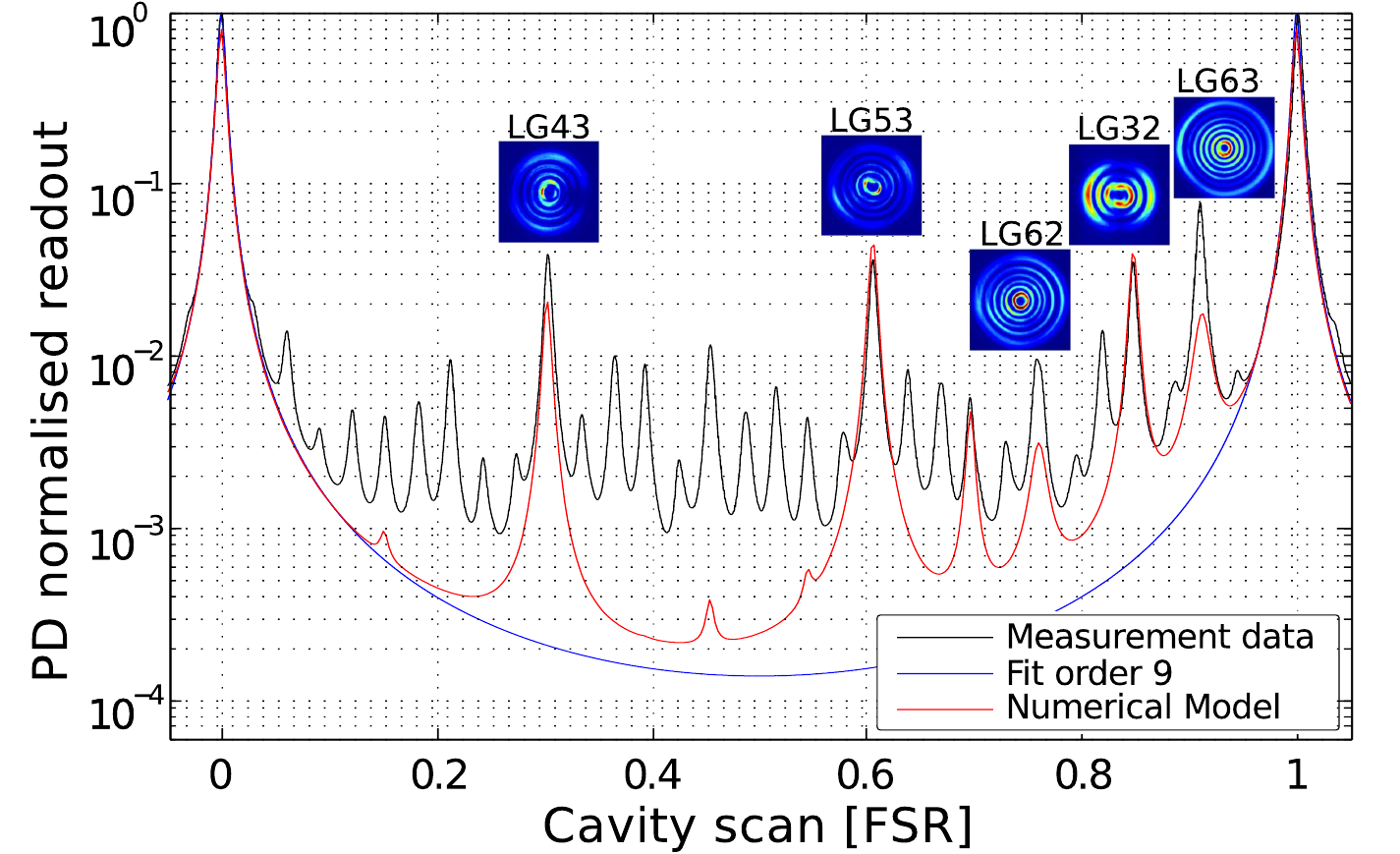}
\caption{\label{fig:LMC_scan} 
Light power transmitted by the MC measured as a function of the cavity length (black line). 
The resonant peaks at 0 and 1 FSR are order 9 modes, nominally \lag. 
The fit to the `\lag' peak measured data is shown for comparison (blue). 
The red curve is derived with a numerical model that assumes the following mode power distribution: 
75\% in LG$_{33}$, 8\% in LG$_{63}$, 4\% in LG$_{43}$, LG$_{53}$ and LG$_{32}$, 1\% in LG$_{62}$.  
The insets show CCD images of these non order 9 modes.
}
\end{center}
\end{figure}

We used measurements of the light transmitted by the MC as a function of the cavity length 
({\it cavity scans}) to investigate the mode content of the 
beam produced by the phase plate, as  
in the example in Fig.\,\ref{fig:LMC_scan}.
The relevant non order 9 modes were first identified via inspection of the CCD images,
then their amplitude, usually a few \% of the total power, and the exact mode content of the overall beam 
could be reproduced with and compared to numerical simulations \cite{ref:FINESSE}. 
On average, the fraction of the beam power in order 9 modes is ($75\pm 5$)\%, 
in agreement with the 
FFT model prediction \cite{fulda:13}. 

\noindent {\bf Results} 
The measurement procedure described above was repeated at progressively increasing input laser power, 
until the maximum available power was injected on the phase plate. 
Increasing the laser power stepwise allowed not only for a prevention 
of damage caused by high powers but also
for identifying the potential rise of power-dependent dynamics and potential 
shortcomings from thermal effects or intra-cavity beam distortions.

\begin{figure}[b]
\begin{center}
\includegraphics[width=8.5cm]{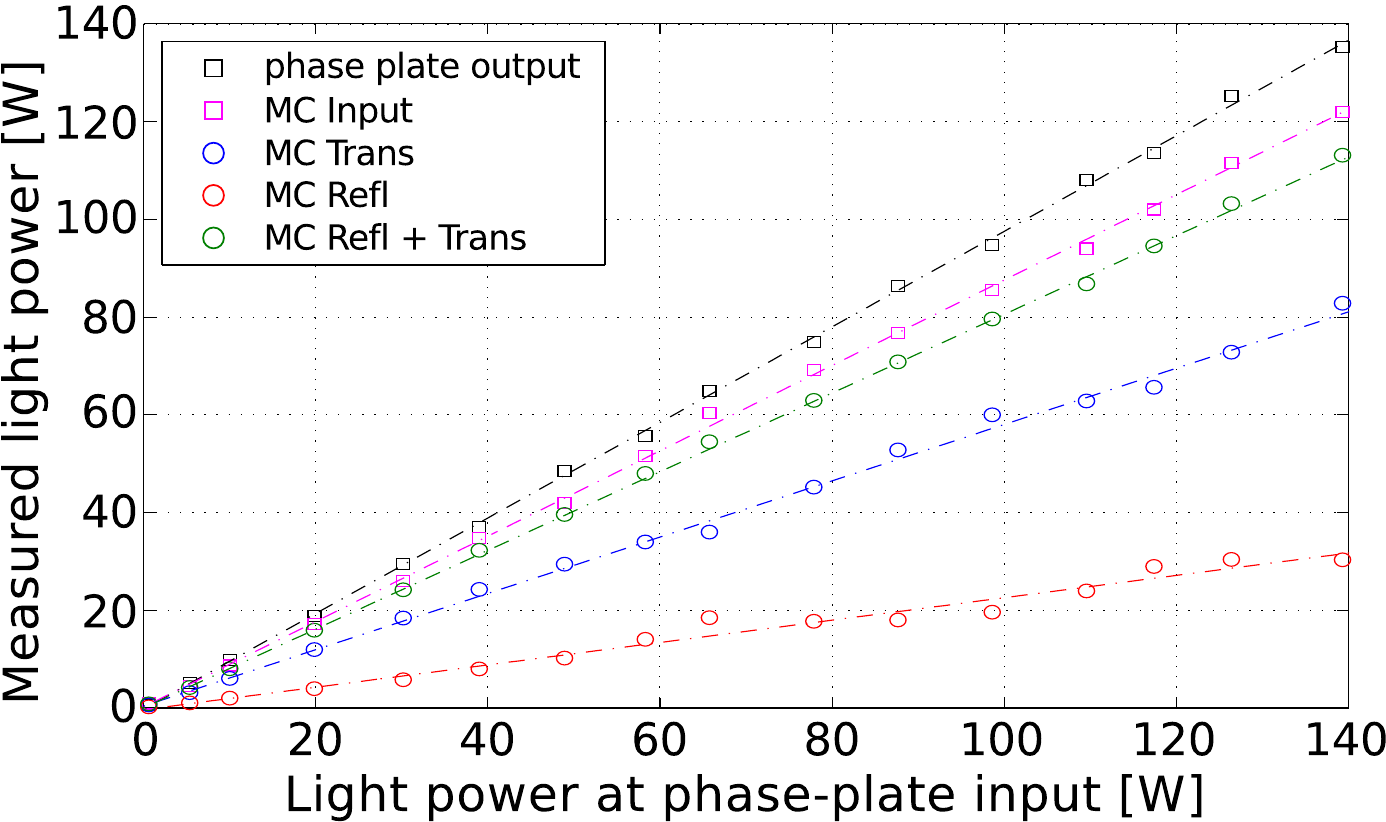}
\caption{\label{fig:powercurves} Measurement of the light power at different locations in the setup as a function 
of the injected laser power.
Statistical uncertainties in the measurement data are smaller than the marker's size and here not reported. 
Systematic errors in the calibration of each power curve is of order 5\%.
}
\end{center}
\end{figure}

\begin{figure}[t]
\begin{center}
\includegraphics[width=8.5cm]{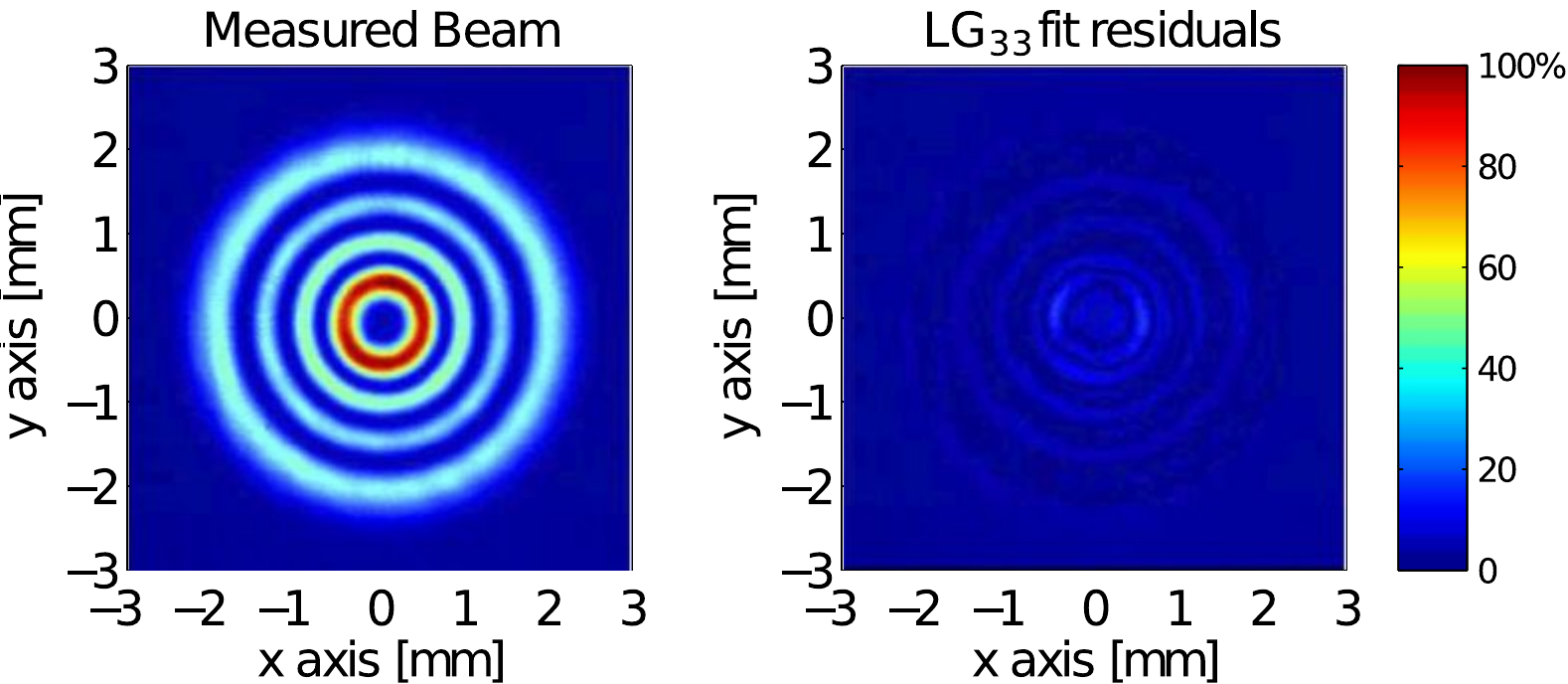}
\caption{\label{fig:lg33_82W} Intensity profile of the 83\,W 
\lag beam transmitted by the MC cavity (left) compared with fit residuals (right). Maps have same units and scale.}
\end{center}
\end{figure}

We show the main results of this experimental 
campaign in Fig.\,\ref{fig:powercurves}, 
where we plot the light power measured at different locations along the setup 
as a function of the incident \lgzero beam power. 
First, the linear response of the power transmitted from the phase plate 
indicates that no effects such as light absorption 
are arising in the phase plate as the power scales up. 
We also plot for completeness the same beam 
when it is propagated to the input of the MC. 
The 7\%  reduction in power is consistent with losses likely 
arising in the intermediate auxiliary optical components and with uncertainties 
in the measurement calibration \footnote{The error in the calibration of each measurement curve 
is of order 5\% and depends on the beam size at the specific measurement position, 
on the auxiliary components utilised in each case and on the instruments themselves.}.
The most notable results in Fig.\,\ref{fig:powercurves} are the measurements of the light 
power reflected and transmitted by the MC when this is resonant to order 9 modes. 
Also in this case the system response is largely linear:
the MC length could be locked to the resonance up to full power, for a maximum 83\,W{\it clean} 
\lag mode transmitted from the MC when a 122\,W {\it raw} LG beam was injected at input. 
To identify potential power-dependent degradations in the mode content of the generated beam, 
cavity scan analyses were made at every laser power level. 
The non order 9 content increased by no more than 5\% at maximum power, 
confirming that expected heating processes are arising in
some component of the beam generation path,
however at a scale which is reasonably small for this type of setup.
Even so, the structure of the \lag output beams did not degrade up to the highest power levels, 
as shown in the example 
in Fig.\,\ref{fig:lg33_82W} where we plot the intensity profile 
of the 83\,W transmitted by the MC.

\begin{figure}[t]
\begin{center}
\includegraphics[width=8.5cm]{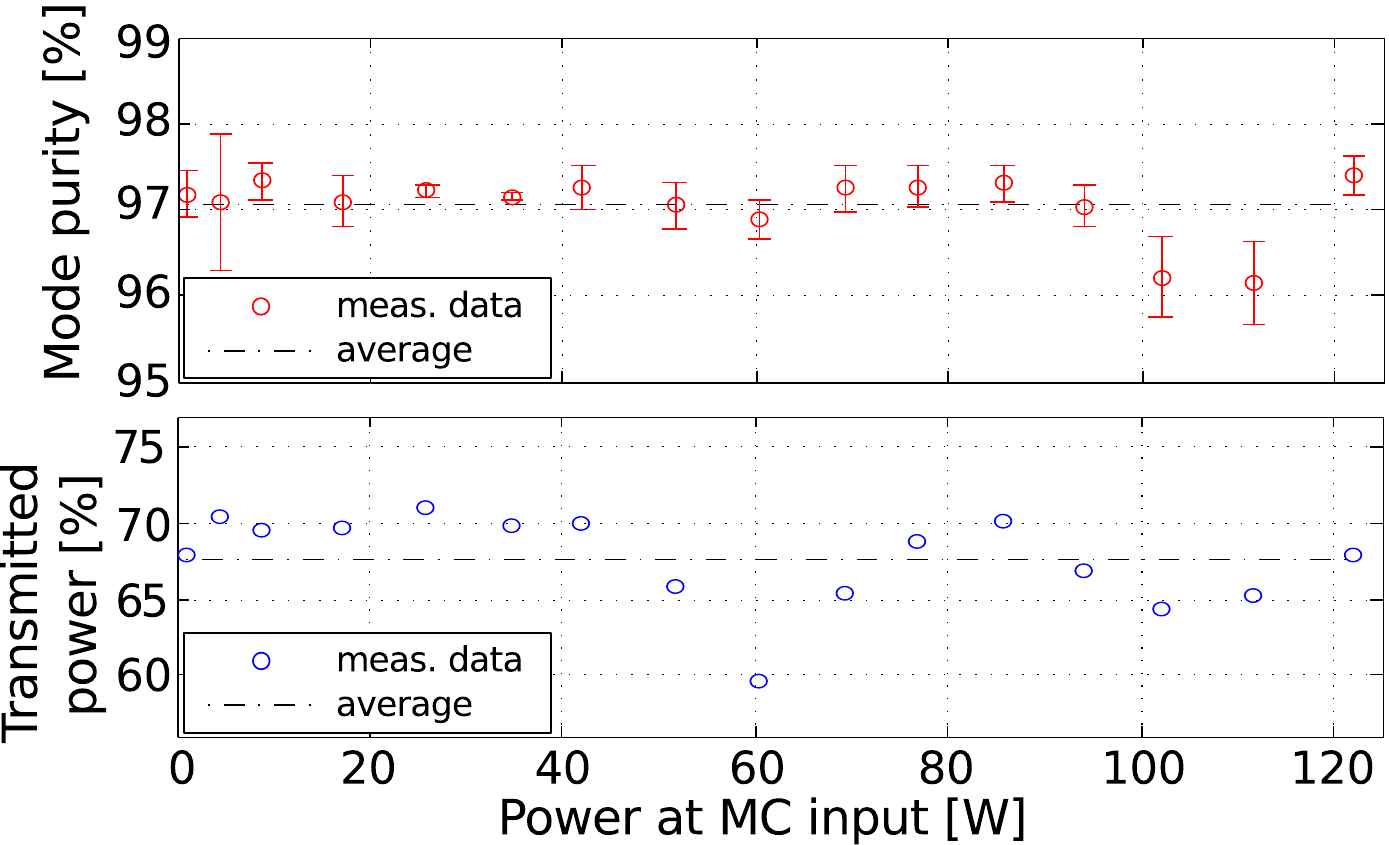}
\caption{\label{fig:overlap} Top: \lag purity of the beam transmitted by the mode cleaner.
Bottom: fraction of the injected light power which is transmitted from the resonant mode cleaner (circles). 
Dashed lines show average values.
Statistical errors are negligible, while 
calibration uncertainty is here of order 7\%.
}
\end{center}
\end{figure}

We assess the purity of the {\it clean} \lag beam as the fraction of power in the beam 
which is in the desired mode and estimate it via the squared inner product 
$\left< {\rm LG}_{33} | \sqrt{\rm{I}_{\rm meas}}\right>^2$ 
between the theoretical \lag amplitude distribution 
and the one measured with the CCD camera, $\sqrt{{\rm I}_{\rm meas}}$
\footnote{We note that since the CCD measures beam intensities, this measurement is in principle degenerate 
for beams with radial index $p=3$ and azimuthal indices $l=\pm3$.}.
Results are shown in Fig.\,\ref{fig:overlap}\,(top) as a function of the correspondent \lag beam power. 
Over the range of investigation, the \lag mode purity is 
above {95\%}, 
and no clear trend or degradation is observed. 
In Fig.\,\ref{fig:overlap}\,(bottom) we show the fraction of the injected 
light power which is transmitted by the resonant MC cavity. 
On average, 68\% is transmitted into a {\it pure} \lag beam,
for a \lag MC cavity throughput about 90\%
Also here, no trend can be observed. 
Taking into account losses in the rest of the apparatus, 
the overall \lgzero to \lag conversion efficiency is 
about 59\%.

\noindent {\bf Summary and conclusions} 
Our experimental investigation into the generation of higher-order LG beams at high laser power 
proved successful: a 83\,W \lag beam with purity 
of order 97\% was obtained 
from a $138$\,W \lgzero laser beam, 
sent through a phase plate and a linear cavity.
To our knowledge at the time of writing, this is  the
highest power ever reported for a higher-order LG beam.
As a by-product, we have also shown that profiling 
of LG beams can be performed at the same 
level of accuracy commonly achieved with \lgzero beams.

The beam generation method seems viable for high power applications.
The system response was mostly linear over the entire range of investigation.
The conversion efficiency, here partly limited by losses in auxiliary optics, 
can be easily improved with an engineered design of the conversion apparatus, 
up to a maximum set by the conversion efficiency of the phase plate design. 
Stability and noise performances were not investigated in this study.

In this work, we have described a method to create a user-defined LG
mode from a highly stable, high-power laser. We have successfully 
demonstrated that this technique creates modes of high purity 
with a good conversion efficiency and is compatible with common setups used 
for the laser pre-stabilisation and injection to GW
interferometers. This is an important step towards 
demonstrating technical readiness of LG modes
for use in high-precision interferometry and in particular
for future GW detectors such as the Einstein Telescope,
as well as for the many other areas of science and technology where LG modes 
have recently found successful application.

{\noindent \bf Acknowledgements} 
This work was funded by 
the ``Science and Technology Facilities Council'' (U.K.)
and the ``Volkswagen Stiftung'' (Germany). 
C.B. acknowledges financial support by
the ``Hannover School of Lasers, Optics and Space-Time Research''. 
This document has LIGO laboratory document number 
LIGO-P1300012-v2.

\end{document}